%% file: nhCO.tex
\documentclass[12pt,preprint]{aastex}

\usepackage{graphicx}
\usepackage{lineno}

\slugcomment{For submission to ApJ}

\shorttitle{Blazars Emission Environment}
\shortauthors{Furniss et al.}

\begin{document}

\title{The Blazar Emission Environment: \\ Insight from Soft X-ray Absorption}


\author{A. Furniss\altaffilmark{1},
M. Fumagalli\altaffilmark{2,3,4},
A. Falcone\altaffilmark{5}, and
D. A. Williams\altaffilmark{1}
}

\altaffiltext{1}{Santa Cruz Institute of Particle Physics and Physics Department, University of California Santa Cruz, 1156 High Street, Santa Cruz, CA 95064, USA}
\altaffiltext{2}{Carnegie Observatories, 813 Santa Barbara Street, Pasadena, CA 91101, USA.}
\altaffiltext{3}{Department of Astrophysics, Princeton University, Princeton, NJ 08544-1001, USA.}
\altaffiltext{4}{Hubble Fellow}
\altaffiltext{5}{Department of Astronomy and Astrophysics, Penn State University, University Park, PA 16802}
\email{afurniss@ucsc.edu}
\email{mfumagalli@obs.carnegiescience.edu}

\begin{abstract}
Collecting experimental insight into the relativistic particle populations and emission mechanisms at work within TeV-emitting blazar jets, which are spatially unresolvable in most bands and have strong beaming factors, is a daunting task.  New observational information has the potential to lead to major strides in understanding the acceleration site parameters.  Detection of molecular  carbon monoxide (CO) in TeV emitting blazars, however, implies the existence of intrinsic gas, a connection often found in photo-dissociated region models and numerical simulations.  The existence of intrinsic gas within a blazar could provide a target photon field for Compton up-scattering of photons to TeV energies by relativistic particles.  We investigate the possible existence of intrinsic gas within the three TeV emitting blazars RGB\,J0710+591, W\,Comae and 1ES\,1959+650 which have measurements or upper limits on molecular CO line luminosity using an independent technique which is based on the spectral analysis of soft X-rays.   Evidence for X-ray absorption by additional gas beyond that measured within the Milky Way is searched for in \textit{Swift} X-ray Telescope (XRT) data between 0.3 and 10 keV.  Without complementary information from another measurement, additional absorption could be misinterpreted as an intrinsically curved X-ray spectrum since both models can frequently fit the soft X-ray data.  After breaking this degeneracy, we do not find evidence for intrinsically curved spectra for any of the three blazars.  Moreover, no evidence for intrinsic gas is evident for RGB\,J0710+591 and W\,Comae, while the 1ES\,1959+650 XRT data support the existence of intrinsic gas with a column density of $\sim$1$\times10^{21}$cm$^{-2}$. 
\end{abstract}


\section{Introduction}
Within the canonical classification of active galactic nuclei (AGN), blazars are a type of active galaxy with a relativistic jet pointed toward the observer.  These galaxies harbor relativistic particle populations with intrinsic emission characteristics which largely remain difficult to characterize.  Blazars produce a non-thermal spectral energy distribution (SED), characterized by two broad peaks in the $\nu$F$_{\nu}$ representation.  The source of the lower-energy peak can be attributed to the synchrotron radiation of relativistic leptons in the presence of a magnetic field.  The mechanism responsible for this relativistic population of leptons (e. g. diffusive parallel shock acceleration, oblique shock acceleration or magnetic reconnection driven flows) remains an open question. 

The origin of the higher-energy SED peak has been attributed to inverse-Compton up-scattering by the relativistic jet leptons of either the synchrotron photons themselves, namely synchrotron self-Compton (SSC) emission, or a photon field external to the jet, namely external Compton (EC) emission (e. g. \cite{dermer, maraschi}).  This external field of photons could arise from emission from a dusty torus, broad line region or some other source within the blazar.  The key feature that distinguishes EC emission from SSC emission, therefore, is the presence of additional gas that can supply an external photon field.  Alternative models attribute the higher energy peak to hadron acceleration leading to proton synchrotron emission and/or pion production accompanied by resulting cascade emission (e. g. \cite{aharonian2002,bednarek,muecke,dermer2012}).

Blazars are sub-classified as flat-spectrum radio-loud quasars if broad emission lines are visible or, otherwise, as BL Lac objects.  BL Lac objects are sub-categorized based on the frequency location of the lower-energy peak, with low-synchrotron-peaked (LSP) BL Lac objects having a peak below $10^{14}$\,Hz, intermediate-synchrotron-peaked (ISP) BL Lac objects peaking between $10^{14}$ and $10^{15}$\,Hz and high-synchrotron-peaked (HSP) BL Lac objects showing a peak above $10^{15}$ Hz \citep{abdoSED}.  Across these different sub-classes, blazars exhibit a continuous variation in their SED.  This has been interpreted as an evolutionary sequence associated with the variation of the diffuse radiation field in the surroundings of the relativistic jet \citep{fossati,bc02,ghisellini}.  Under this paradigm, LSP blazars are expected to have an appreciable external radiation field which facilitates effective cooling and favors a EC broadband representation while HSPs lack a source of external radiation and would be described with an SSC emission scenario. 

Measurements which can probe for the existence of intrinsic gas which might produce an external radiation field for Compton up-scattering, however, are challenging for BL Lac objects . The bright, non-thermal continuum emitted by the blazar jet makes the discrimination between SSC and EC emission mechanisms difficult.   Broadband modeling of BL Lacs is a common method for the investigation of the non-thermal emission mechanism at work within a blazar jet (e. g. \cite{mrk501,mrk421,pks2155}).  These models utilize various input parameters which describe the non-thermal emission, including the particle injection parameter, $q$, which also represents the index of the relativistic electron energy distribution in the absence of significant cooling processes which can deform the power-law emission.  The relativistic particles are directly responsible for the synchrotron emission which makes up the lower energy SED peak.  Since the spectrum of synchrotron radiation follows from the energy distribution of the emitting particles, the highest energy particles can be directly probed through X-ray observations.    More specifically, the observed radiation spectrum can be translated into the spectrum of relativistic particles emitting that synchrotron radiation. 

Two functional forms are commonly adopted to describe the observed synchrotron spectrum: a power-law model and a log-parabolic model.  A power-law synchrotron spectrum with photon index $\alpha$ originates from a particle power-law energy spectrum with index $q=2\alpha+1$ \citep{rl}, where $q$ is equal to the index of the input particle distribution in the absence of a cooled high energy tail.  Similarly, a log-parabolic radiation spectrum can be produced by a log-parabolic distribution of relativistic particles \citep{paggi} and can be interpreted as evidence for a statistical acceleration process \citep[e.g.]{tramacere2007,massaro2006,massaro2004,costamante2001,tagliaferri2003}.  This log-parabolic emission can also be interpreted as representing a power-law particle distribution with a cooled high energy tail, producing an intrinsically curved X-ray spectrum.  An accurate determination of the spectral shape of the underlying population is key in the elucidation of the relativistic acceleration mechanism at work within a blazar jet. Fundamentally different acceleration methods produce varying particle spectra, and intrinsic spectral curvature is a possible signature of statistical acceleration.

Regardless of the intrinsic spectral shape of the synchrotron emission, the non-thermal X-ray radiation from any extragalactic X-ray emitter undergoes absorption by the interstellar medium of our own Galaxy.  The effective energy-dependent cross section results from the absorption of photons by intervening elements such as (but not limited to) C, N and O, an effect extending up to 10 keV.  These absorption cross sections have been calculated per neutral hydrogen column density (N$_{\rm HI}$) as a function of energy in the 0.03-10 keV range \citep{xraycrosssection} and can be used to estimate the amount of absorption extragalactic X-ray emission undergoes because of the intervening gas within the Milky Way.

Both an absorbed power law and an absorbed log parabola have been shown to produce acceptable fits to blazar spectral data between 0.3 and 10 keV.  The hydrogen column density that is used to scale the abundance of the additional absorbing elements is typically fixed to the Galactic value as measured by the LAB survey \citep{kalberla}.  In practice, these fits do not allow for the presence of additional gas in the blazar host galaxy.   There have, however, been a limited number of blazars which, when fit with an absorbed power-law model with a free column density, show a fitted value in excess to that measured in the Milky Way (e. g. \cite{rxj0648,es1959}).  However, some ambiguity in the interpretation of the data remain.  For these instances, the use of an absorbed log-parabolic model with a fixed Galactic N$_{\rm HI}$ column density allows for an intrinsically curved spectrum and can provide an improved fit to the X-ray data without requiring additional absorption within the blazar emission region.

We argue that an apparently curved spectrum observed in TeV BL Lacertae objects may, in some cases, be partially or entirely due to intrinsic blazar column density beyond that measured within our own Galaxy, similarly to what was found in \cite{tavecchio2002} for a flat-spectrum radio quasar.   Detection of additional X-ray absorption exceeding that expected from the Milky Way, therefore, might provide an experimental probe of the emission mechanism at work within TeV blazars (e. g. SSC or EC emission).  To investigate the degeneracy between intrinsic curvature in the X-ray spectrum and additional absorption from gas in the blazar host galaxy, we utilize independent observations of the gas content in the surroundings of the blazars. 

Recent millimetric observations of CO have probed the molecular gas content in three TeV-emitting blazars \citep{fumagalli}.   A common result of photo-dissociation region (PDR) models and numerical simulations is that CO molecules form in regions where the UV radiation has been shielded by gas and dust, (e. g. \cite{tielenn,krumholz,gloverclark}).  The detection of molecular gas in a galaxy is therefore a signpost of additional intervening material in the surroundings of the blazar, similar to the interstellar gas that absorbs soft X-rays within our own Galaxy.   While the exact conversion between the observed CO column density and the hydrogen column density (both in atomic and molecular phase)  would require a detailed knowledge of the chemical and physical state of the gas as well as of the impinging radiation field, it is reasonable to expect  that the presence of CO is associated with a hydrogen column density $>$10 M$_{\odot}$pc$^{-2}$ ($\sim1\times10^{21}$cm$^{-2}$). Similarly, in normal conditions it would be surprising to find regions in which the N$_{\rm HI}$ column density greatly exceeds 10 M$_{\odot}$pc$^{-2}$ without a molecular phase. 

Millimetric observations that are sufficiently deep to map CO in blazars have thus far only been completed for three sources, namely RGB\,J0710+591, W\,Comae and 1ES\,1959+650.  With the understanding that a sample of three blazars is far too limited to draw general conclusions, we investigate the relation between the presence of molecular gas and evidence for intrinsic column density in blazars as well as intrinsic X-ray spectral curvature, as inferred from spectral fitting. 

In this work we show that the possibility of combining independent diagnostics at different frequencies has immediate implications for the physical interpretation of emission processes at work within TeV-emitting blazar jets.  In Section 2 we discuss the \textit{Swift} XRT observations and data reduction which allow for the investigation of  additional absorption by intrinsic gas.  In Section 3 we summarize the results of the spectral analysis for each source, addressing the possibility for both intrinsic absorption and spectral curvature.  Finally, we discuss our results in a broader context with regard to blazar non-thermal emission and the implications for the blazar evolutionary theory.

\section{Observations and Data Reduction}
The data set includes all \textit{Swift} X-ray Telescope (XRT) exposures of RGB\,J0710+591, W\,Comae, and 1ES\,1959+650 since the launch of the instrument in November 2004 \citep{gehrels}.  The summary of each exposure used can be found in Tables 1-4.   The XRT is a grazing incidence Wolter-I telescope which focuses X-rays between 0.2 and 10 keV onto a 100 cm$^2$ CCD \citep{burrows}.  The data were analyzed using the \texttt{HEASoft} package v6.12.  Event files were cleaned and calibrated with standard criteria with the \textit{xrtpipeline} task, using only events between 0.3 and 10 keV having a grade 0-12 for windowed timing (wt) mode and 0-2 for photon counting (pc) mode.  Additionally, events having energy between 0.4 and 0.7 keV were removed from the analysis due to the effects of the oxygen edge at 0.54 keV\footnote{\texttt{heasarc.nasa.gove/docs/swift/proposals/swift\_responses.html}}.  More details on the \textit{Swift} XRT effective area can be found in \cite{evans}.

For windowed timing mode, rectangular source regions with length and width of 45 and 8 pixels were used for spectral analysis.  Background regions were extracted from similarly sized rectangular regions of nearby source-free sky, aligned along the windowed timing one dimensional stream in sky coordinates.  

For photon counting mode, circular source regions of radius 20 pixels and centered on the source were used along with similarly sized background regions of nearby source-free sky.  If a source showed a count rate greater than 0.5 counts per sec, pileup was accounted for by using an annular source region with inner radius of 2 pixels and outer radius 20 pixels to remove the pile-up contamination in the inner part of the point spread function.  For 1ES\,1959+650, a source showing consistently high count rates between 3 and 12 counts per second, we only include windowed timing exposures lasting more than 0.5 ks.

The response matrices\footnote{Windowed timing analysis used \texttt{swxwt0to2s6\_20070901v012.rmf} and photon counting analysis used \texttt{swpc0to12s6\_20070901v011.rmf}.} available from CALDB were used to create ancillary response files with \textit{xrtmkarf}.  For single observations with sufficient statistics to allow spectral fitting, source spectra were binned to require 30 counts per bin.  Single exposures providing count levels too low to allow for the bin minimum required for accurate $\chi^2$ minimization fitting were first summed with \textit{xselect}, with the exposure files summed with \textit{ximage}, before being binned to require 30 counts per bin.  This summation was only necessary for the low count levels found for W\,Comae.  Summing all photon counting mode exposures of W\,Comae provided 176 bins with more than 30 counts.  It is noted that the summation of these exposures introduced a possibility for summing different spectral states.

\section{Spectral Analysis and Results}
The spectral analysis was performed with \texttt{XSPEC}\footnote{\tt http://heasarc.nasa.gov/docs/software/lheasoft/xanadu/xspec/XspecManual.pdf} Version 12.7.1 and a suite of in-house IDL routines.  We fit the data with two spectral models.  The first is an absorbed power-law model \textit{(wabs(powerlaw)} in XSPEC), with an additional exponent to represent additional neutral hydrogen column density beyond the Milky Way

\begin{equation}
 F(E)_{PL}= Ke^{-(N^{\rm MW}_{\rm HI}+N^{\rm INT}_{\rm HI})\sigma(E)} (E)^{-\alpha} ,
 \end{equation}
 
\noindent referred to as the PL model for the remainder of this work, where $\alpha$ is the spectral slope, K is the normalization parameter and $\sigma(E)$ is the non-Thompson energy dependent photoelectric cross section from \cite{xraycrosssection}.  N$^{\rm MW}_{\rm HI}$ is the neutral hydrogen column density within the Milky Way, as measured by the LAB survey\footnote{\cite{kalberla}, \texttt{www.astro.uni\-bonn.de/hisurvey/profile/}}.  The LAB-measured neutral hydrogen densities for each blazar are 4.16$\times10^{20}$cm$^{-2}$, 1.97$\times10^{20}$cm$^{-2}$ and 1.00$\times10^{21}$cm$^{-2}$ for RGB\,J0710+591, W\,Comae and 1ES\,1959+650, respectively.  N$^{\rm INT}_{\rm HI}$ is the blazar neutral hydrogen column density, both in units of cm$^{-2}$.  We note that this analysis does not allow differentiation between blazar-intrinsic and intervening gas residing along the line of sight to the blazar.  With the possibility for galaxy clustering, the local environment around the blazars may contain an overdensity of galaxies that can contribute to the observed absorption.  Any additional column density found through this analysis is assumed to be in the vicinity of the blazar.

The LAB survey consists of observations of 21-cm emission from Galactic neutral hydrogen over the entire sky.  The LAB Survey is a sensitive Milky Way neutral hydrogen survey, with extensive coverage both spatially and kinematically.  The survey merges the Leiden/Dwingeloo Survey (LDS: \cite{hartman}) of the sky north of $-30$ degrees declination with the Instituto Argentino de Radioastronom'a Survey (IAR: \cite{arnal,bajaja}) of the sky south of $-25$ degrees declination. Uncertainties in N$^{\rm MW}_{\rm HI}$ column densities, as reported by the LAB survey, are between 2 and 3\%, with $\sim$1\% from scale uncertainty and $\sim$ 1\% from uncertainties in the correction for stray radiation.   

The second spectral model applied to the data is an absorbed log-parabola \textit{(wabs(logpar)} in XSPEC) with an additional exponent to represent additional neutral hydrogen column density beyond the Milky Way, referred to as the LP model for the remainder of this work. This model has been suggested to better represent TeV-detected blazar X-ray spectra (e. g. \cite{massaro, tramacere2007}).   This model allows the spectral index to vary as a function of energy, according to the equation 

\begin{equation}
F(E)_{LP}=Ke^{-(N^{\rm MW}_{\rm HI}+N^{\rm INT}_{\rm HI})\sigma(E)}(E)^{-(\alpha+\beta {\rm log}(E))},
 \end{equation}

\noindent with a normalization factor K, photoelectric cross section and hydrogen column densities similar to those described for the PL model above.  

These models were fit to the XRT data of the three blazars with both N$^{\rm INT}_{\rm HI}$ set to zero and with N$^{\rm INT}_{\rm HI}$ allowed to vary, while requiring it to remain equal to or greater than zero.  
The best fits for each of the models are shown in Figure 1 for each of the blazars.  These fits are completed for the Observation ID 0003156006 for RGB\,J0710+591, the summed exposure of W\,Comae and Observation ID 00035025004 for 1ES\,1959+650.  For RGB\,J 0710+591 and 1ES\,1959+650, the fits are shown for the exposure with the highest statistics and are meant to be representative of the fitting results.  The fitted parameters from Figure 1 are shown in bold in Tables 1, 3 and 4.

The parameter spaces for the spectral models as applied to the data are illustrated by reduced $\chi^2$ contours in Figures 2-4.  The N$^{\rm INT}_{\rm HI}$ column density is allowed to vary when producing these contours, except for left hand column of Figure 3, when illustrating the fitted $\alpha$ and $\beta$ parameters resulting for the LP model with and without a  N$^{\rm INT}_{\rm HI}$ column density.    These contours show parameter constraints (or lack thereof) during the spectral $\chi^2$ fitting procedures.  The white dotted, dashed and dash-dotted lines represent the one, two and three sigma confidence contours on the joint distribution of parameters.

\subsection{RGB\,J0710+591}
The spectral models were applied to each of the 13 \textit{Swift} XRT observations of RGB\,J0710+591 and are summarized in Table 1.  When fit with both PL and LP models which allow for N$^{\rm INT}_{\rm HI}$ column density, there is no significant evidence for the existence of additional gas, e. g. N$^{\rm INT}_{\rm HI}$ parameters are consistent with zero.  These fits provide parameter values equivalent to the N$^{\rm INT}_{\rm HI}$-lacking fits.  Due to this redundancy, the resulting parameters are only quoted for fits where N$^{\rm INT}_{\rm HI}$ is not included.   The exposures provide comparable fit qualities for the N$^{\rm MW}_{\rm HI}$ only and N$^{\rm INT}_{\rm HI}$ inclusive models.

As shown in Figure 3 for the LP model, the fitted $\alpha$ and $\beta$ parameters do not change significantly when allowing for a non-zero N$^{\rm INT}_{\rm HI}$ column density.  Additionally, the lack of constraint on the N$^{\rm INT}_{\rm HI}$ column density with respect to these parameters is evident in the top panel of Figure 4.   When fitting the data with a LP model, negligible curvature ($\beta$) is found for a majority of the observations.  This $\beta \sim 0$ result indicates that the PL model is sufficient to describe the X-ray emission from this blazar. 

There are a few instances where significant negative curvature is found from the LP fitting.  This unexpected result is likely due to covariance between the $\alpha$ and $\beta$ parameters ( i. e. moving diagonally in the Figure 3 parameter spaces).  One can see that in each instance where a negative curvature parameter is found, the fitted index $\alpha$ softens to values of $\sim$2 or less.   Due to the unlikely possibility to produce a concave distribution of relativistic particles by traditional acceleration mechanisms, a more physical result might be found by fitting a log-parabolic model with $\beta>0$ limitations.  

\subsection{W\,Comae}
The PL and LP spectral models were applied to the summed photon counting mode exposures of W\,Comae.  The 73 observations summed to 1.3$\times10^5$ seconds and are summarized in Table 2, with the fitting results in Table 3.  As illustrated by the confidence contours in the middle panels of Figures 2 and 4, application of the PL and LP models which allow for non-zero N$^{\rm INT}_{\rm HI}$ column density result in N$^{\rm INT}_{\rm HI}$ parameters consistent with zero, showing no significant evidence for an intervening gas within the blazar and providing parameter values equivalent to the fixed N$^{\rm MW}_{\rm HI}$ fits.  

Similarly to the LP model result for RGB\,J0710+591, the fitted $\alpha$ and $\beta$ parameters do not change significantly when including N$^{\rm INT}_{\rm HI}$.  The negative curvature parameter $\beta$ found when applying the LP model to the summed W\,Comae spectra may indicate a correlation in the $\alpha$ and $\beta$ parameters, similarly as the trend presented in the RGB\,J0710+591 data.  This specific case of negative curvature might also be a result of summing multiple exposures.  Investigation of spectral effects due to summing were completed by fitting the summed exposure while ignoring bins between 1.5 and 2.2 keV in order to remove possible pollution from the Si K$\alpha$ edge.  Removing these bins, however, did not change the spectral results.   

Spectral effects introduced through the summation of different spectral states and/or flux levels may also introduce distorted spectral features such as concavity (i.e. the summation of a bright soft state with a low hard state).  The previously observed broadband variability of blazars at all wavebands and timescales probed supports this possibility.  X-ray spectral variability of W\,Comae during a flare, however, shows the source to harden during periods of elevated flux \citep{wcom}.  We take this behavior as evidence that the concavity of the summed spectrum is not likely due to the summation of bright soft and low hard states.   The 73 observations show the count rate of W\,Comae to vary between 0.05 and 3.8 counts per second.  Details of X-ray variability of BL Lac objects are described in \cite{bc02}. 

If the resulting $\chi^2$ values are used at face value to determine which model best represents W\,Comae, one would conclude that the curved LP model is the most appropriate.  However, the convex curvature of the radiation spectrum is not easily described by standard acceleration mechanisms and so we favor a PL model over a LP model for W\,Comae.

\subsection{1ES\,1959+650}
When fitting both the PL and LP models to 1ES\,1959+650, a significant  N$^{\rm INT}_{\rm HI}$ column density is derived (see Tables 3 and 4 for a summary of all results).  
The $\alpha$ and $\beta$ parameter for the LP model can be seen to change significantly when N$^{\rm INT}_{\rm HI}$ is included in the fit (Figure 3).  The $\beta$ parameter becomes consistent with zero when N$^{\rm INT}_{\rm HI}$ is included in the fit, indicating that if the possibility for intrinsic gas in the blazar emission environment is allowed, the particle population can be represented with an uncurved power-law spectrum.  Evidence for additional absorption exceeding that by the Milky Way was also found for power-law fits of XMM X-ray observations of 1ES\,1959+650 in \cite{perlman}, although the additional column density was interpreted as evidence for a curved log-parabolic spectrum.

The variation of the PL-fitted blazar column density between observations is illustrated in the light curve in Figure 5, showing the PL N$^{\rm INT}_{\rm HI}$ column densities in comparison to the large flux variations exhibited by the blazar ($\chi^2=35089$ for 61 degrees of freedom).  The N$^{\rm INT}_{\rm HI}$ values are less variable ($\chi^2=320$ for 61 degrees of freedom) with an average of 8$\times10^{20}$cm$^{-2}$, nearly equal to the Galactic contribution of 1.0$\times10^{21}$cm$^{-2}$, as measured by the LAB survey.   If the variability of the column density is real, it would suggest that the gas is local to the non-thermal emission region.  X-ray variability of AGN is commonly observed (e.g. \cite{mushotzky,vagnetti}) and has been attributed to the variation of intrinsic absorption and covering fraction of the galaxy \citep{abrassart, wachter}.  As a geometrically selected subtype of jetted AGN, variability of column density within a blazar is possible.  However, the variability of column density for 1ES\,1959+650 shown in Figure 5 might result from the underestimation of the N$^{\rm INT}_{\rm HI}$ errors as they do not account for correlations between fitting parameters.

\section{Discussion and Conclusion}
The 0.3-10 keV spectral analysis of the three blazars which have molecular CO line luminosity measurements and upper limits support a possible connection between blazar CO line luminosity and hydrogen column density, providing a possible method to directly probe the relativistic particle energy distribution.  The two sources RGB\,J0710+591 and W\,Comae, both of which are shown to exhibit no CO line luminosity to the limits probed in \cite{fumagalli}, also lack evidence for X-ray absorption by intrinsic gas and could be sufficiently described with an uncurved power-law.   

The blazar 1ES\,1959+650 shows significant curvature when fit with a LP model.  The LP fit qualities are comparable to those found when an intrinsic column density is included for the PL model.  The blazar has been shown to contain a significant level of molecular CO.  Relying on the connection between molecular CO and hydrogen column density form PDR, the corresponding column density is $\sim1\times10^{21}$cm$^{-2}$, or more, is in agreement with the typical value of $0.8\times10^{21}$cm$^{-2}$ we obtain in the fits to the X-ray spectra.  We have therefore broken the degeneracy between the LP and the N$^{\rm INT}_{\rm HI}$-inclusive PL models, concluding that this source is likely to be exhibiting an intrinsically uncurved X-ray spectrum with intrinsic gas leading to the observed spectra.

We note that in the LP spectral analysis of RGB\,J0710+591 and W\,Comae, instances of significantly negative spectral curvature result from a covariance relation between the $\alpha$ and $\beta$ parameters.  Additional investigation to this relation is beyond the scope of this paper.  However, future application of the LP model to similar X-ray spectra might result in more physical models if completed with a limit to positive curvature ($\beta>0$) within the LP model framework, as convex spectra are difficult to produce under traditional non-thermal emission scenarios, although not impossible.

The uncurved X-ray spectra found for these sources suggest that the particle population responsible for the synchrotron emission is similarly uncurved, supporting a non-statistical (energy-independent) acceleration process at work within the blazar.  The information about the shape of the particle energy distribution can be put to use in broadband models of non-thermal emission from TeV-blazars, which require the input of the population distribution shape and slope.  If the relation between CO line luminosity and the evidence for intrinsic X-ray absorption holds for other blazars, similar observations can be used to place previously unavailable model constraints on other X-ray bright blazars.  By extension, the use of X-ray spectral analysis in search of evidence for additional intrinsic absorption can provide a hint as to how bright the CO line luminosity for certain blazars might be.

The apparent lack of intrinsic gas in the blazar RGB\,J0710+591 also supports past broadband modeling of the non-thermal emission, which shows acceptable representation from a SSC emission scenario, indicating a clean emission environment \citep{rgbj0710}  Similar modeling for W\,Comae, however, shows a preference for an EC emission scenario, relying on an external photon field which might arise from intrinsic gas \citep{wcom}.   The lack of intrinsic gas and acceptable SSC modeling is in agreement with the blazar evolutionary theory for the HSP RGB\,J0710+591, but the lack of evidence for intrinsic gas within W\,Comae is in contradiction to what the blazar evolution would imply for the emission environment of the ISP.

The evidence for intrinsic gas within the HSP 1ES\,1959+650 is also contrary to the blazar evolutionary theory, which predicts a clean emission environment for the blazar.  This result does, however, agree with the previously detected variability patterns displayed by this blazar which are not easily supported by clean SSC emission scenarios (e. g. an ``orphan" flare event, reported in \cite{krawczynski}).  Moreover, this variability pattern has been described by the presence of intrinsic dilute gas \citep{boettcher05}, as suggested by the independent molecular CO measurements and the N$^{\rm INT}_{\rm HI}$ inferred from soft X-ray absorption.  Other observations (e. g. \cite{meyer, krawczynski}) have also found evidence contradictory to the current blazar evolutionary theory.

In addition, we note that the presence of an intrinsic column density comparable to the Galactic column density ($\sim$1$\times10^{21}$cm$^{-2}$) for 1ES\,1959+650 would introduce more than one magnitude of extinction for $\lambda < 3000$ \AA\  (assuming standard gas-to-dust ratio and Milky Way extinction law).  Non-thermal UV emission (e. g. as seen in \cite{tramacere2007}) is suppressed by additional column density, deforming the observed SED from that which is emitted.  Evidence for intrinsic column density within the blazar environment would support the use of an additional extinction-correction to the observed UV flux values beyond that required by the Milky Way to more accurately represent the intrinsically emitted SED.

In summary, the evidence for intrinsic gas in the HSP 1ES\,1959+650 and the lack of it in the ISP W\,Comae do not directly align with the blazar evolutionary paradigm.  However, the millimetric and X-ray observations for these sources do support a possible connection between the existence of molecular CO and intrinsic gas, maintaining the possibility for a rare opportunity to collect experimental insight into the emission environment and mechanism at work within blazar jets.

\acknowledgments
The authors are grateful to the Swift Team for their efforts to provide these observations.  Support for A. Furniss and D. Williams was provided by NASA grant NNX12AJ12G and NSF grant PHY09-70134.  Support for M. Fumagalli was provided by NASA through Hubble Fellowship grant HF-51305.01-A awarded by the Space Telescope Science Institute, which is operated by the Association of Universities for Research in Astronomy Inc., for NASA, under contract NAS 5-26555.   A. Falcone acknowledges support from NASA grant NNX10AU14G.

{\it Facilities:} \facility{Swift (XRT)}.

\begin{figure}
\epsscale{0.8}
\plotone{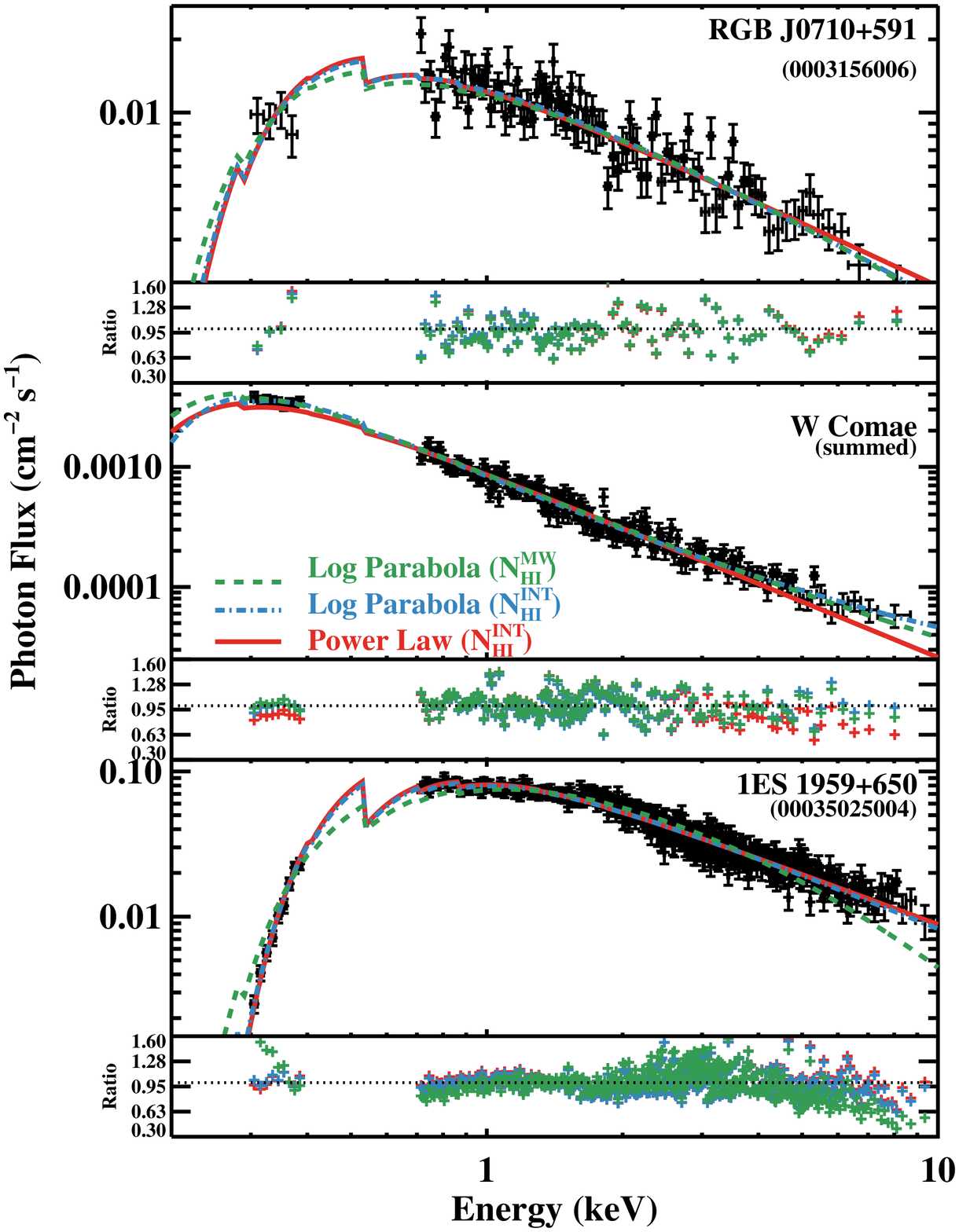}
\caption{The best fit PL and LP models for the RGB\,J0710+591 Observation 0003156006, the W\,Comae summed exposure and the 1ES\,1959+650 Observation 00035025004.  The ratios of the data to models are shown in the bottom portion of each panel, with the dotted line drawn at a ratio of one. The PL model is shown with fitted N$^{\rm INT}_{\rm HI}$ column density (red solid line), while the LP models are shown for both N$^{\rm INT}_{\rm HI}$ lacking (green dashed line) and N$^{\rm INT}_{\rm HI}$ contributing (blue dash-dotted line) column densities.  The feature around $\sim$0.5 keV arises from the onset of the oxygen interaction cross-section.  The fitted parameters are summarized in Tables 1, 3 and 4 in bold.}\label{fig1}
\end{figure}

\begin{figure}
\epsscale{0.8}
\plotone{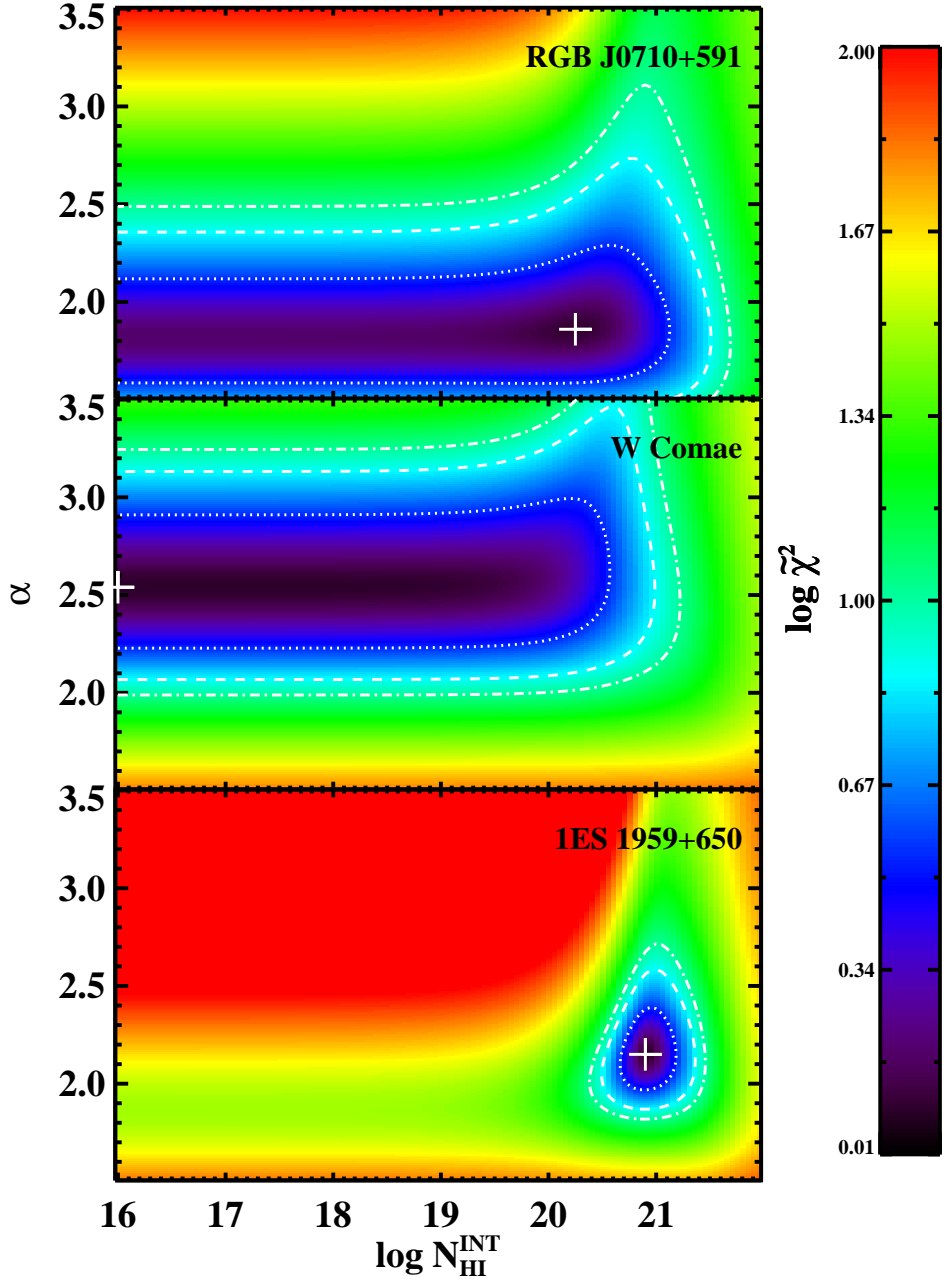}
\caption{The reduced $\chi^2$ contours for the PL model for the RGB\,J0710+591 Observation 0003156006, the W\,Comae summed exposure, and the 1ES\,1959+650 Observation 00035025004.  The white dotted, dashed and dash-dotted lines represent the one, two and three sigma confidence contours on the joint distribution of parameters.  The N$^{\rm INT}_{\rm HI}$ parameter for RGB\,J0710+591 and W\,Comae (top and middle panels) are shown to be unconstrained, showing that the data do not favor any additional absorption from intrinsic gas within the blazar and beyond that of the Milky Way.  The minimum of the fit for RGB\,J0710+591, however, lies within the defined parameter space, while it does not for the summed W\,Comae data.}\label{fig2}
\end{figure}

\begin{figure}
\epsscale{0.8}
\plotone{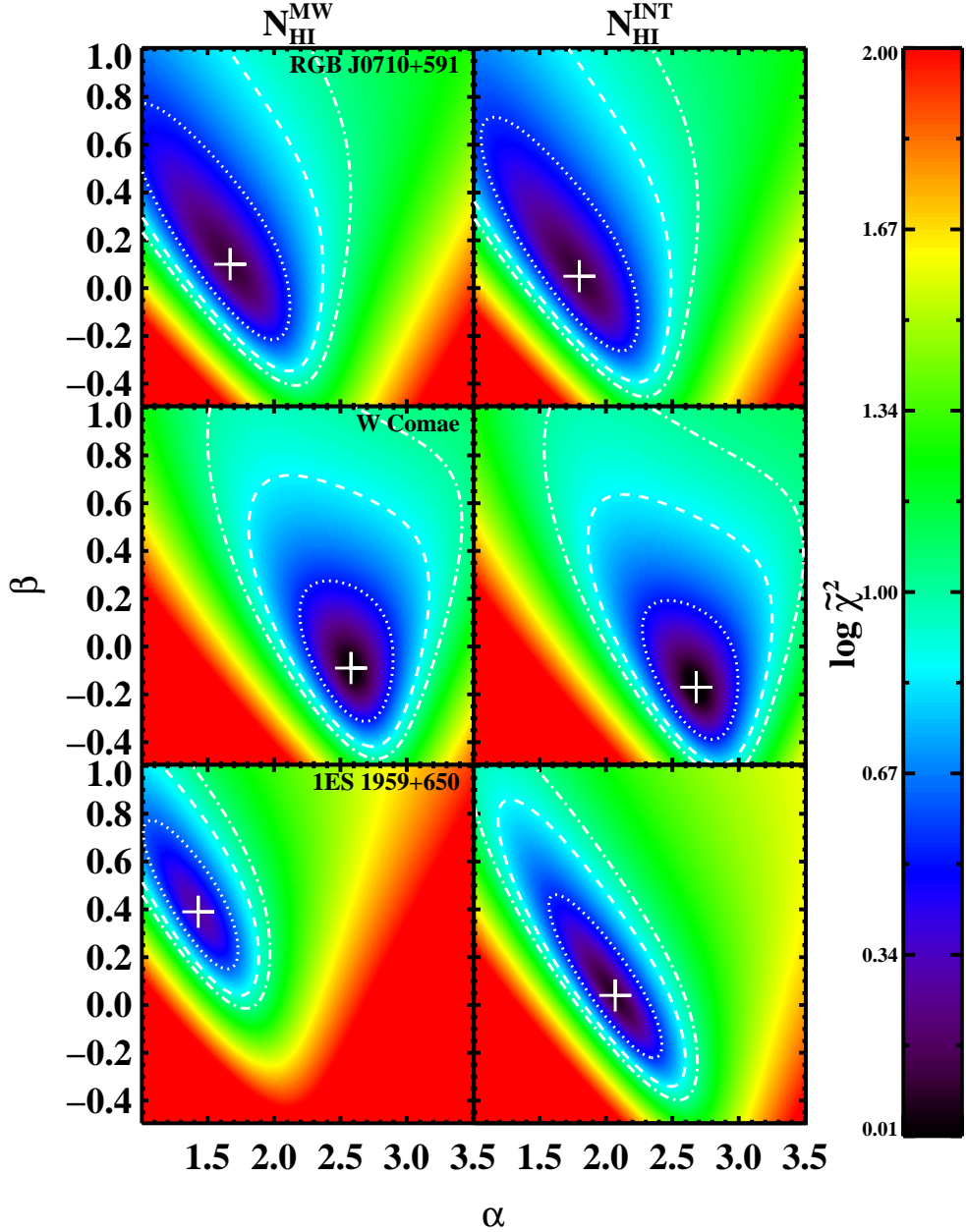}
\caption{Reduced $\chi^2$ contours for the LP model for the RGB\,J0710+591 Observation 0003156006, the W\,Comae summed exposure and the 1ES\,1959+650 Observation 00035025004 for N$^{\rm INT}_{\rm HI}$ lacking (left) and inclusive (right) column density fits.  The contours are similarly represented, as described in Figure 2.  The index $\alpha$ and curvature $\beta$ parameters are not seen to significantly change for free and fixed N$_{\rm HI}$ fits for RGB\,J0710+591 and W\,Comae.  For 1ES\,1959+650, the $\alpha$ and $\beta$ parameters change from 1.5 and 0.4 to 2.1 and 0, respectively, when N$^{\rm INT}_{\rm HI}$ is included in the fit.  This change suggests that if N$^{\rm INT}_{\rm HI}$ column density is included, no intrinsic curvature in the spectrum of 1ES\,1959+650 is necessary to match the observed spectrum. Note that the x- and y-axes are interchanged in these plots compared to Figure 2.}\label{fig3}
\end{figure}

\begin{figure}
\epsscale{0.8}
\plotone{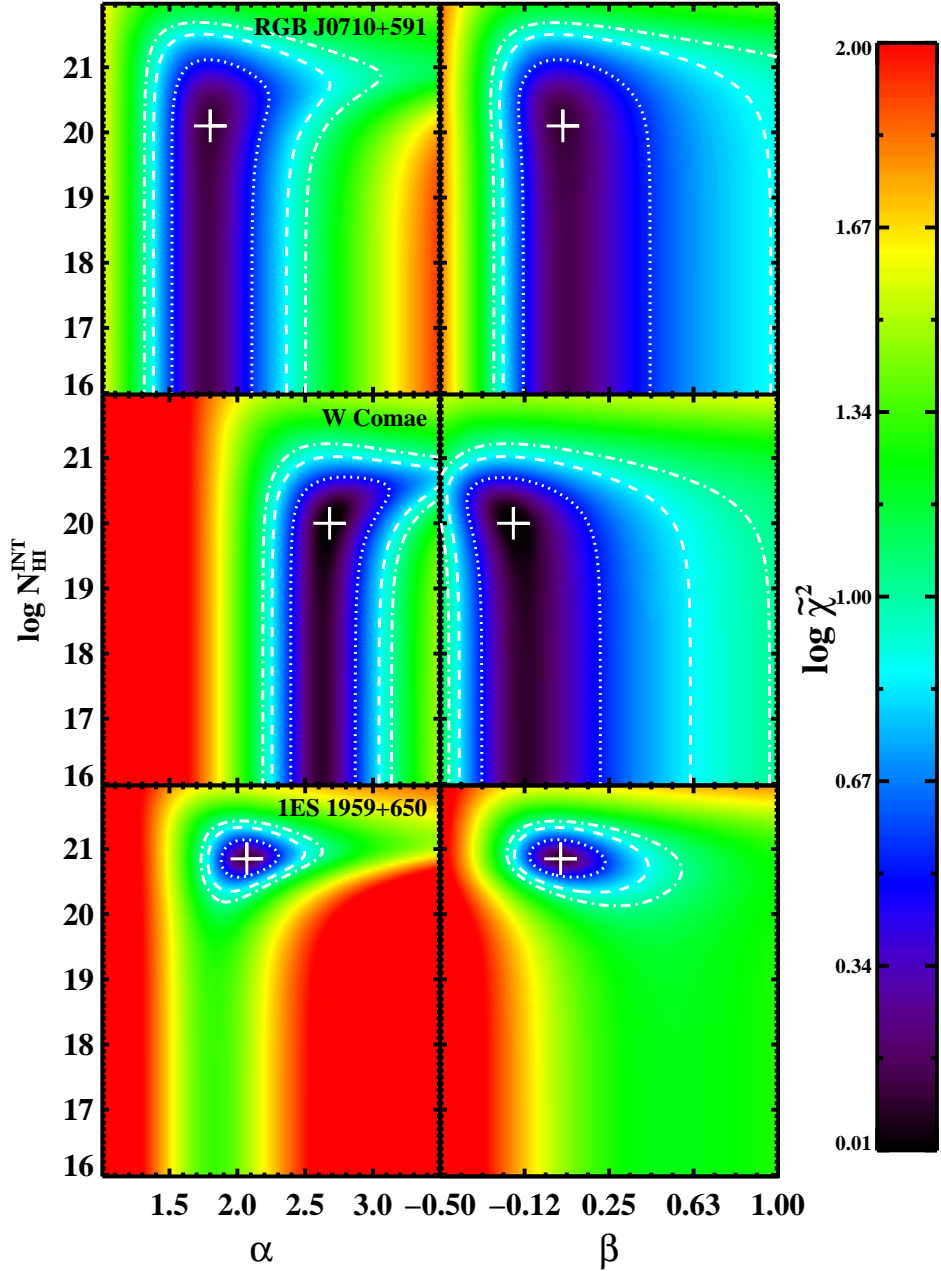}
\caption{Reduced $\chi^2$ contours for the LP model for the RGB\,J0710+591 Observation 0003156006, the W\,Comae summed exposure and the 1ES\,1959+650 Observation 00035025004 for fitted N$^{\rm INT}_{\rm HI}$ column density.  The contours are similarly represented, as described in Figure 2.  The column densities of RGB\,J0710+591 and W\,Comae are seen to be unconstrained, while they are well defined for 1ES\,1959+650, showing an intrinsic column density of order 1$\times10^{21}$cm$^{-2}$, in addition to the 1$\times10^{21}$cm$^{-2}$ as measured by the LAB Galactic N$_{\rm HI}$ survey.  }\label{fig4}
\end{figure}

\begin{figure}
\includegraphics[angle=90,scale=0.6]{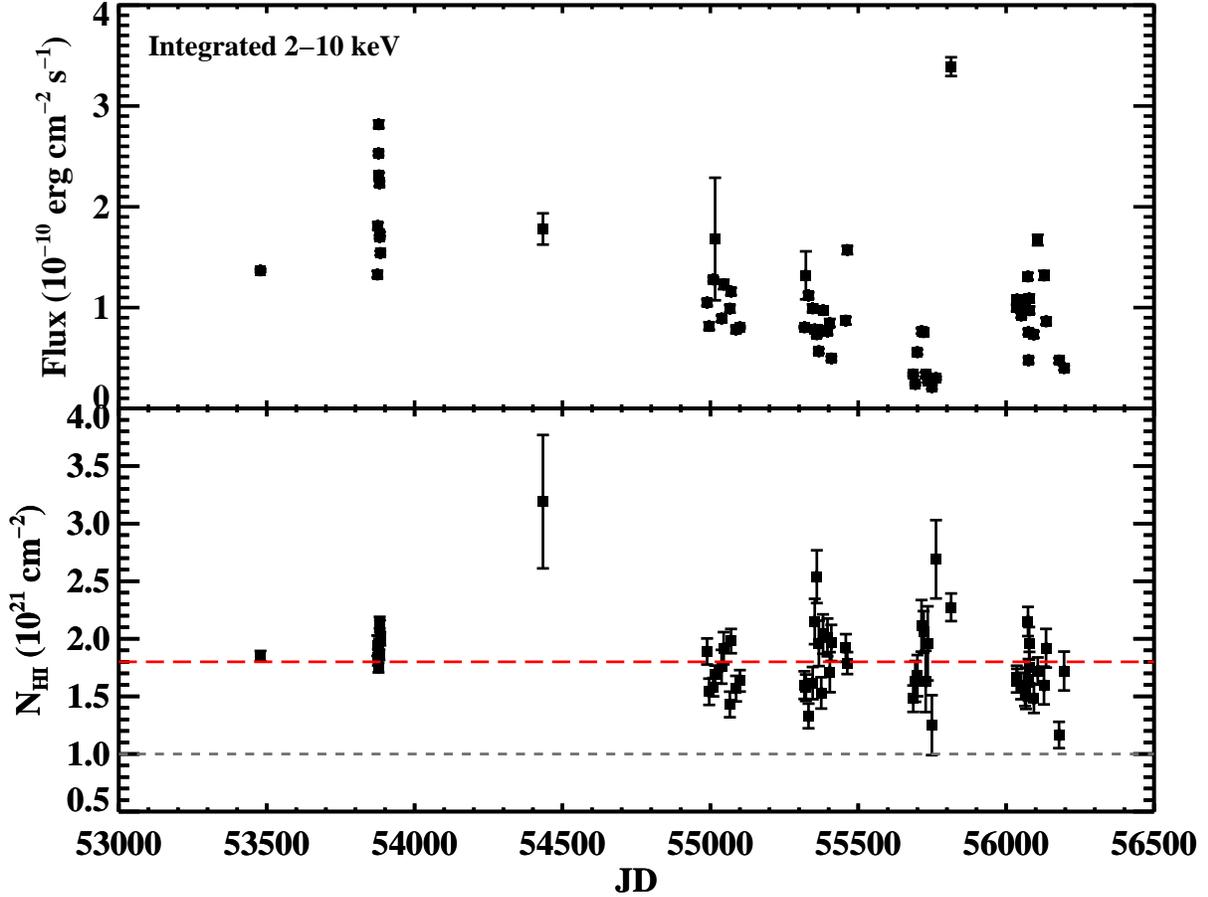}
\caption{A light curve of the 2$-$10 keV flux and fitted N$^{\rm INT}_{\rm HI}$ column density for the 61 windowed timing observations of 1ES\,1959+650.  The flux is seen to vary widely, with a fit to a constant resulting in a $\chi^2$ of 35089 for 61 degrees of freedom. The N$^{\rm INT}_{\rm HI}$ column density is also variable, but shows a smaller $\chi^2$ value of 320 when fit with a constant, with an average of 0.8$\times10^{21}$cm$^{-2}$ (denoted by the red dashed line), nearly equal to the value N$^{\rm MW}_{\rm HI}$ of 1.0$\times10^{21}$cm$^{-2}$ as reported by the LAB survey (denoted by the grey dotted line).  This additional 0.8$\times10^{21}$cm$^{-2}$ column density beyond that of the LAB value is in significant excess of the 1-2\% error reported for the LAB survey measurements.   The slight variability of the N$^{\rm INT}_{\rm HI}$ column density is not immediately expected and, if real, suggests that the column density is very close to the non-thermal emission region.  However, the errors shown here do not account for slight correlations between other fitted PL parameters.}\label{fig5}
\end{figure}

\rotate
\input{nHCOTableRGB.tex}
\input{WComExposureTable.tex}
\input{nHCOTableWCom.tex}
\input{nHCOTable1ESB.tex}
\input{nHCOTable1959LP.tex}
\end{document}

%% file: nHCOTableRGB.tex
\begin{deluxetable}{ccccccccccccc}
\tabletypesize{\scriptsize}
\tablecaption{Summary of spectral analysis RGB J0710+591 for absorbed PL and absorbed LP.  When the possibility of N$^{\rm INT}_{\rm HI}$ was included in the fits, the results were consistent with zero, providing similar fit parameters as found when N$^{\rm INT}_{\rm HI}$ was set to zero.    For the fits here, N$^{\rm MW}_{\rm HI}$ was fixed to 4.16$\times10^{20}$cm$^{-2}$ as measured by \cite{kalberla} and N$^{\rm INT}_{\rm HI}$ is set to zero.  Observation ID 0003156006 is shown in bold and represents the best fit values for the models shown in the top panel of Figure 1. }
\tablewidth{0pt}
\tablehead{
  \colhead{Observation}&  \colhead{Date}&  \colhead{Mode} &\colhead{Exposure}&  \colhead{Spectral} &     \colhead{}& \colhead{2-10 keV} &  \colhead{}&  \colhead{}&   \colhead{}& \colhead{}\\
  \colhead{ID}&  \colhead{}& \colhead{}& \colhead{}&  \colhead{Data} &    \colhead{PL}&  \colhead{Flux} & \colhead{LP}&  \colhead{LP}&  \colhead{PL}&  \colhead{LP}\\
    \colhead{}&  \colhead{[MJD]}&\colhead{}&  \colhead{[ks]}&  \colhead{Counts} &  \colhead{$\alpha$}& \colhead{[$\times10^{-11}$ ergs cm$^{-2}$s$^{-1}$]} &  \colhead{$\alpha$}&  \colhead{$\beta$}&  \colhead{$\chi^2$/dof}& \colhead{$\chi^2$/dof}\\  
  
}
 \startdata
0003156001&54882&pc&3.19&3348 &1.76$\pm$0.03&3.5$\pm$0.1&1.68$\pm$0.06&0.1$\pm$0.1&81.82/82&80.14/81\\
0003156002&54887&pc&1.97& 1491&1.83$\pm$0.05&3.3$\pm$0.2&1.6$\pm$0.1&0.3$\pm$0.2&31.13/37&29.3/36\\
0003156003&54888&pc&1.84&1978 &1.81$\pm$0.04&3.6$\pm$0.1&1.71$\pm$0.08&0.2$\pm$0.1&62.55/48&61.01/47\\
0003156004&54889&wt&1.97&4231 &1.77$\pm$0.03&5.1$\pm$0.2&1.78$\pm$0.05&-0.02$\pm$0.07&97.05/93&97/92\\
0003156005&54890&wt&1.91& 2824&1.79$\pm$0.03&4.6$\pm$0.2&1.69$\pm$0.07&0.2$\pm$0.1&69.16/66&65.97/65\\
\textbf{0003156006}&\textbf{54891}&\textbf{wt}&\textbf{2.19}&\textbf{4735}&\textbf{1.81$\pm$0.03}&4.6$\pm$0.2&\textbf{1.73$\pm$0.05}&\textbf{0.10$\pm$0.08}&\textbf{128.09/100}&\textbf{122.37/99}\\
0003156007&54892&wt&1.97& 4278&1.78$\pm$0.03&4.9$\pm$0.2&1.76$\pm$0.05&0.05$\pm$0.08&104.55/95&104.09/94\\
0003156008&55975&wt&1.25&1685 &1.97$\pm$0.05&3.2$\pm$0.2&2.09$\pm$0.07&-0.3$\pm$0.1&37.64/39&33.42/38\\
0003156009&55978&wt&1.09& 1375&2.07$\pm$0.06&1.9$\pm0.2$&2.18$\pm$0.09&-0.2$\pm$0.2&21.16/31&19.53/30\\
0003156010&55985&wt&0.97&1286 &1.93$\pm$0.06&2.5$\pm$0.2&2.2$\pm$0.1&-0.6$\pm$0.1&41.59/29&28.09/28\\
0003156011&55998&wt&1.15& 1477&1.84$\pm$0.06&2.6$\pm$0.2&1.97$\pm$0.09&-0.3$\pm$0.1&31.44/34&28.68/33\\
0003156012&56002&wt&0.86&1154 &1.91$\pm$0.06&2.6$\pm$0.2&2.1$\pm$0.1&-0.3$\pm$0.2&31.78/26&29.08/25\\
0003156013&56007&wt&1.03&1304 &2.15$\pm$0.06&1.7$\pm$0.2&2.44$\pm$0.06&-0.8$\pm$0.1&72.35/28&43.7/27\\
 \enddata
\end{deluxetable}

%% file: WComExposureTable.tex
\begin{deluxetable}{cccc}
\tabletypesize{\scriptsize}
\tablecaption{Summary of \textit{Swift} XRT exposures of W\,Comae, summed for spectral analysis.}
\tablewidth{0pt}
\tablehead{
  \colhead{Observation}&  \colhead{Date}&  \colhead{Exposure}& \colhead{Source}\\
  \colhead{ID}&  \colhead{[MJD]}&  \colhead{[ks]} & \colhead{Counts}\\  }
 \startdata
 0003092001&54223&2.45&194\\
0003092002&54225&4.56&503\\
0003092003&54226&3.70&403\\
00031160001&54539&4.38&458\\
00031160003&54541&2.15&173\\
00031160004&54553&1.68&436\\
00031160005&54554&1.63&168\\
00031219001&54624&8.98&4390\\
00031219002&54626&5.09&525\\
00035018001&53672&1.40&40\\
00031219002&53627&9.89&706\\
00031219003&53720&8.58&1050\\
00031219004&54495&2.26&94\\
00031219005&54499&4.56&122\\
00031219006&54528&2.06&327\\
00031219007&54536&1.86&80\\
00031219009&54589&1.72&264\\
00031219010&54828&1.52&74\\
00031219011&54835&1.03&110\\
00031219012&54842&0.98&60\\
00031219013&54844&0.94&92\\
00031219014&54856&0.66&209\\
00031219015&54863&1.12&189\\
00031219016&54870&1.31&167\\
00031219017&54877&0.96&112\\
00031219018&54884&1.35&372\\
00031219019&54891&1.54&138\\
00031219020&54898&1.19&64\\
00031219021&54912&1.59&96\\
00031219022&54919&0.87&39\\
00031219024&54933&0.98&48\\
00031219025&55175&4.81&182\\
00031219026&55230&1.02&60\\
00031219027&55237&1.04&21\\
00031219028&55245&1.18&45\\
00031219029&55251&1.23&14\\
00031219031&55265&0.88&25\\
00031219032&55272&0.94&174\\
00031219033&55279&1.09&34\\
00031219034&55286&0.73&270\\
00031219035&55293&0.93&549\\
00031219036&55300&1.16&893\\
00031219038&55314&1.35&367\\
00031219039&55335&0.95&297\\
00031219040&55356&1.97&668\\
00031219041&55599&0.97&267\\
00031219042&55605&0.89&370\\
00031219043&55612&1.12&475\\
00031219044&55626&1.09&579\\
00031219045&55636&0.96&235\\
00031219046&55641&1.19&290\\
00031219047&55648&1.14&353\\
00031219048&55661&0.95&374\\
00031219049&55669&1.20&1725\\
00031219050&55675&1.14&539\\
00031219051&55691&1.09&328\\
00031219053&55710&0.94&418\\
00031219054&55727&2.33&622\\
00031219055&55728&2.10&572\\
00031219056&55960&1.04&3143\\
00031219057&55967&1.16&358\\
00031219058&55974&1.33&414\\
00031219059&55981&1.03&289\\
00031219060&55988&1.47&454\\
00031219061&56002&0.92&338\\
00031219062&56010&1.37&661\\
00031219063&56023&1.01&334\\
00031219064&56037&0.88&294\\
00031219065&56042&0.92&289\\
00031219066&56053&0.99&283\\
00031219067&56058&1.05&449\\
00031219068&56065&1.02&766\\
00031219069&56072&1.05&364\\
 \enddata
\end{deluxetable}

%% file: nHCOTableWCom.tex
\begin{deluxetable}{cccccccccc}
\tabletypesize{\scriptsize}
\tablecaption{Summary of summed exposure spectral analysis for W\,Comae for an absorbed PL and LP model. The count rate for W\,Comae was low enough to require the summation of all photon counting mode exposures before grouping to 30 counts per bin.  When  N$^{\rm INT}_{\rm HI}$ was included in the fits, the results were consistent with zero and provided redundant fit parameters as found when N$^{\rm INT}_{\rm HI}$ was set to zero.  For the fits quoted here, the N$^{\rm MW}_{\rm HI}$ column density is fixed to 1.97$\times10^{20}$cm$^{-2}$, as found in the LAB survey \citep{kalberla} and N$^{\rm INT}_{\rm HI}$ is set to zero.  The best fit models are shown for the summed exposure in the middle panel of Figure 1.}
\tablewidth{0pt}
\tablehead{
  \colhead{Exposure}&  \colhead{Spectral} &  \colhead{PL}& \colhead{2-10 keV Flux}&   \colhead{LP}&  \colhead{LP}&  \colhead{PL}& \colhead{LP}\\
  \colhead{[ks]}&  \colhead{Counts} &  \colhead{$\alpha$}& [$\times10^{-12}$ ergs cm$^{-2}$s$^{-1}$]  &  \colhead{$\alpha$}&  \colhead{$\beta$}&  \colhead{$\chi^2$/dof}&  \colhead{$\chi^2$/dof}\\  }
 \startdata
\textbf{132}&\textbf{9806}&\textbf{2.51$\pm$0.02}& \textbf{1.10$\pm$0.02}&\textbf{2.56$\pm$0.02}&\textbf{-0.13$\pm$0.04}&\textbf{191.94/174}&\textbf{163.52/173}\\
 \enddata
\end{deluxetable}

%% file: nHCOTable1ESB.tex
\begin{deluxetable}{cccccccccc}
\tabletypesize{\scriptsize}
\tablecaption{Summary of PL spectral analysis for 1ES\,1959+650.  Only windowed timing exposures of more than 500 seconds are included in this analysis.  The Galactic N$_{\rm HI}$ column density was fixed to 1$\times10^{21}$cm$^{-2}$, as measured by \cite{kalberla} with a $\sim$3\% error.  Observation ID 00035025004 is shown in bold and represents the best fit values for the models shown in the bottom panel of Figure 1.}
\tablewidth{0pt}
\tablehead{
  \colhead{Observation}&  \colhead{Date}&  \colhead{Exposure} &  \colhead{Fixed}&  \colhead{PL}&  \colhead{Free}&  \colhead{2-10 keV Flux}&  \colhead{Spectral}&  \colhead{Fixed}& \colhead{Free}\\
  \colhead{ID}&  \colhead{}&  \colhead{}&  \colhead{PL}&  \colhead{N$^{\rm MW}_{\rm HI}$+N$^{\rm INT}_{\rm HI}$}&  \colhead{PL}&  \colhead{$\times10^{-11}$}&  \colhead{Counts} &   \colhead{PL}& \colhead{PL}\\
    \colhead{}&  \colhead{[MJD]}&  \colhead{[ks]} &  \colhead{$\alpha$}&  \colhead{[$\times10^{21}$cm$^{-2}$]}&  \colhead{$\alpha$}&  \colhead{[ergs cm$^{-2}$s$^{-1}$]}& \colhead{}&    \colhead{$\chi^2$/dof}& \colhead{$\chi^2$/dof} }
 \startdata
00035025001&53479&4.43&2.102$\pm$0.008&1.852$\pm$0.004&2.40$\pm$0.02&13.7$\pm$0.1&32782&1019.0/330&380.3/329\\
00035025002&53874&1.43&2.01$\pm$0.02&1.943$\pm$0.009&2.32$\pm$0.03&13.3$\pm$0.3&9270&426.3/192&223.1/191\\
00035025003&53876&1.99&1.97$\pm$0.01&1.846$\pm$0.007&2.22$\pm$0.02&18.1$\pm$0.3&14911&510.3/251&289.5/250\\
\textbf{00035025004}&\textbf{53878}&\textbf{5.35}&\textbf{1.91$\pm$0.01}&\textbf{1.753$\pm$0.004}&\textbf{2.13$\pm$0.01}&\textbf{25.3$\pm$0.2}&\textbf{44642}&\textbf{1019.0/417}&\textbf{500.8/416}\\
00035025005&53879&2.30&1.90$\pm$0.01&1.960$\pm$0.006&2.19$\pm$0.02&28.2$\pm$0.4&26758&804.9/328&384.1/327\\
00035025006&53880&4.38&1.95$\pm$0.008&1.957$\pm$0.005&2.22$\pm$0.01&23.1$\pm$0.3&44306&1211.8/401&483.8/400\\
00035025007&53881&4.37&1.975$\pm$0.008&1.865$\pm$0.004&2.23$\pm$0.01&22.3$\pm$0.3&43562&1119.9/396&480.4/397\\
00035025008&53882&4.28&2.078$\pm$0.008&2.147$\pm$0.005&2.43$\pm$0.02&17.0$\pm$0.2&36841&1430.8/343&458.0/342\\
00035025009&53883&4.41&2.067$\pm$0.008&2.117$\pm$0.005&2.41$\pm$0.02&17.3$\pm$0.2&35858&1265.9/344&429.3/343\\
00035025010&53884&3.28&2.12$\pm$0.01&2.001$\pm$0.006&2.44$\pm$0.02&15.4$\pm$0.2&26439&914.7/244&383.6/243\\
00035025016&54434&1.28&2.07$\pm$0.06&3.192$\pm$0.058&2.5$\pm$0.1&17.8$\pm$1.6&1268&56.0/37&40.2/36\\
00035025027&54989&0.99&2.26$\pm$0.02&1.899$\pm$0.011&2.56$\pm$0.04&10.5$\pm$0.3&6399&267.0/150&165.9/149\\
00035025028&54996&1.0&2.33$\pm$0.03&1.542$\pm$0.011&2.52$\pm$0.05&8.18$\pm$0.4&4749&147.1/120&116.3/119\\
00035025032&55010&1.55&2.17$\pm$0.02&1.587$\pm$0.008&2.36$\pm$0.03&12.8$\pm$0.2&9745&239.2/195&166.1/194\\
00035025034&55016&1.64&2.11$\pm$0.02&1.693$\pm$0.008&2.32$\pm$0.03&16.8$\pm$6.1&12162&329.9/219&210.1/218\\
00035025037&55038&0.60&2.28$\pm$0.03&1.762$\pm$0.015&2.56$\pm$0.06&8.91$\pm$0.3&3331&126.2/84&87.6/83\\
00035025038&55045&0.65&2.19$\pm$0.03&1.927$\pm$0.014&2.46$\pm$0.05&12.3$\pm$0.5&4687&179.2/120&116.6/119\\
00035025041&55066&1.20&2.17$\pm$0.02&1.439$\pm$0.011&2.31$\pm$0.04&9.88$\pm$0.4&5488&144.4/139&124.3/138\\
00035025042&55070&1.23&2.15$\pm$0.02&1.989$\pm$0.011&2.48$\pm$0.04&11.6$\pm$0.4&7914&330.5/175&177.0/174\\
00035025043&55087&0.96&2.27$\pm$0.03&1.575$\pm$0.011&2.47$\pm$0.05&78.1$\pm$0.3&4693&140.4/120&106.1/119\\
00035025044&55100&6.12&2.256$\pm$0.009&1.646$\pm$0.004&2.47$\pm$0.02&80.6$\pm$0.1&30252&619.9/303&320.2/302\\
00035025045&55318&1.29&2.07$\pm$0.02&1.606$\pm$0.012&2.24$\pm$0.04&80.6$\pm$0.1&4836&180.6/122&151.3/121\\
00035025046&55323&1.09&2.03$\pm$0.02&1.584$\pm$0.012&2.18$\pm$0.04&13.2$\pm$2.4&6433&163.0/159&129.2/158\\
00035025047&55332&1.02&2.07$\pm$0.02&1.333$\pm$0.011&2.17$\pm$0.04&11.2$\pm$0.3&5188&124.0/132&113.9/131\\
00035025048&55346&1.12&2.07$\pm$0.03&1.622$\pm$0.014&2.24$\pm$0.04&9.94$\pm$0.4&5081&122.1/130&94.3/129\\
00035025049&55353&1.08&2.05$\pm$0.03&2.152$\pm$0.020&2.32$\pm$0.05&7.83$\pm$0.3&4252&154.0/107&95.8/106\\
00035025050&55360&1.07&2.13$\pm$0.03&2.548$\pm$0.023&2.52$\pm$0.06&7.37$\pm$0.3&4026&206.2/100&98.5/99\\
00035025051&55367&1.01&2.16$\pm$0.03&1.967$\pm$0.020&2.43$\pm$0.06&5.68$\pm$0.2&3326&103.1/82&62.8/81\\
00035025052&55375&1.16&2.16$\pm$0.03&1.536$\pm$0.014&2.32$\pm$0.05&7.77$\pm$0.3&4708&182.2/121&161.2/120\\
00035025053&55381&1.16&2.03$\pm$0.02&2.046$\pm$0.017&2.29$\pm$0.05&9.76$\pm$0.3&5703&227.1/145&165.2/144\\
00035025054&55396&1.11&2.14$\pm$0.03&2.019$\pm$0.016&2.41$\pm$0.05&7.64$\pm$0.3&4644&167.6/118&104.7/117\\
00035025056&55404&0.77&2.23$\pm$0.03&1.710$\pm$0.017&2.44$\pm$0.06&8.48$\pm$0.4&3273&129.5/83&104.9/82\\
00035025057&55409&1.02&2.34$\pm$0.03&1.961$\pm$0.016&2.68$\pm$0.06&4.94$\pm$0.2&3628&137.1/93&73.6/92\\
00035025058&55458&1.26&2.17$\pm$0.02&1.921$\pm$0.012&2.46$\pm$0.04&8.69$\pm$0.4&6132&242.4/147&143.1/146\\
00035025059&55463&1.24&2.01$\pm$0.02&1.798$\pm$0.010&2.24$\pm$0.03&15.7$\pm$0.4&9237&313.8/193&205.7/192\\
00035025062&55686&1.19&2.50$\pm$0.03&1.483$\pm$0.012&2.69$\pm$0.05&3.36$\pm$0.2&3247&115.0/81&92.8/80\\
00035025063&55693&1.25&2.48$\pm$0.04&1.630$\pm$0.018&2.70$\pm$0.07&2.43$\pm$0.1&2438&91.4/66&72.9/65\\
00035025064&55700&0.94&2.31$\pm$0.03&1.689$\pm$0.018&2.50$\pm$0.06&5.59$\pm$0.2&3361&109.9/83&90.6/82\\
00035025066&55714&0.95&2.09$\pm$0.03&2.111$\pm$0.022&2.35$\pm$0.06&7.68$\pm$0.3&3804&127.1/94&88.2/93\\
00035025067&55721&0.98&2.22$\pm$0.03&2.060$\pm$0.018&2.53$\pm$0.06&7.58$\pm$0.3&3625&151.6/93&94.7/92\\
00035025068&55728&1.0&2.28$\pm$0.04&1.632$\pm$0.027&2.46$\pm$0.09&3.35$\pm$0.2&2001&63.2/53&54.9/52\\
00035025069&55735&1.17&2.29$\pm$0.04&1.968$\pm$0.032&2.54$\pm$0.09&2.74$\pm$0.2&2073&71.3/56&54.6/55\\
00035025071&55749&1.00&2.36$\pm$0.06&1.255$\pm$0.026&2.4$\pm$0.1&2.12$\pm$0.2&1442&39.5/40&38.5/39\\
00035025072&55763&1.05&2.32$\pm$0.04&2.696$\pm$0.034&2.77$\pm$0.09&3.01$\pm$0.2&2364&127.8/63&74.1/62\\
00035025074&55814&1.41&2.10$\pm$0.02&2.276$\pm$0.012&2.42$\pm$0.03&33.9$\pm$0.9&11257&442.4/216&233.1/215\\
00035025075&56036&0.98&2.44$\pm$0.03&1.652$\pm$0.012&2.67$\pm$0.05&9.97$\pm$0.3&5504&156.7/130&112.4/129\\
00035025076&56037&1.52&2.36$\pm$0.02&1.678$\pm$0.007&2.60$\pm$0.03&10.8$\pm$0.2&11140&337.4/195&220.7/194\\
00035025077&56051&0.99&2.34$\pm$0.02&1.592$\pm$0.012&2.52$\pm$0.04&9.23$\pm$0.2&6044&154.6/142&117.5/141\\
00035025078&56064&1.06&2.24$\pm$0.02&1.525$\pm$0.011&2.41$\pm$0.04&10.8$\pm$0.3&5500&161.6/135&130.2/134\\
00035025079&56067&0.69&2.23$\pm$0.03&1.574$\pm$0.018&2.38$\pm$0.06&9.79$\pm$0.3&3841&116.6/94&103.4/93\\
00035025080&56074&1.27&2.10$\pm$0.02&2.150$\pm$0.013&2.40$\pm$0.05&13.1$\pm$0.3&9006&370.5/187&232.1/186\\
00035025081&56075&1.10&2.28$\pm$0.02&1.631$\pm$0.012&2.48$\pm$0.04&7.59$\pm$0.3&5230&145.3/127&108.2/126\\
00035025082&56076&0.92&2.40$\pm$0.03&1.643$\pm$0.018&2.60$\pm$0.06&4.77$\pm$0.3&3207&105.4/82&85.2/81\\
00035025083&56078&1.04&2.15$\pm$0.02&1.966$\pm$0.014&2.42$\pm$0.04&10.9$\pm$0.4&6148&214.9/147&138.3/146\\
00035025084&56079&1.52&2.37$\pm$0.02&1.759$\pm$0.014&2.44$\pm$0.04&9.75$\pm$0.3&7082&210.1/161&167.7/160\\
00035025085&56094&0.99&2.40$\pm$0.03&1.488$\pm$0.012&2.57$\pm$0.05&7.32$\pm$0.3&4448&155.1/113&135.8/112\\
00035025086&56107&1.01&2.09$\pm$0.02&1.722$\pm$0.012&2.28$\pm$0.04&16.7$\pm$0.5&6083&212.7/155&159.3/154\\
00035025087&56128&1.01&2.16$\pm$0.03&1.590$\pm$0.016&2.32$\pm$0.05&13.2$\pm$0.5&3916&84.0/97&66.3/96\\
00035025088&56135&1.18&2.27$\pm$0.03&1.927$\pm$0.017&2.54$\pm$0.05&8.65$\pm$0.3&4892&153.7/126&103.1/125\\
00035025089&56179&0.93&2.48$\pm$0.03&1.167$\pm$0.011&2.53$\pm$0.05&4.74$\pm$0.3&3142&98.2/77&96.1/76\\
00035025091&56196&1.52&2.38$\pm$0.03&1.721$\pm$0.017&2.59$\pm$0.06&4.01$\pm$0.2&3934&119.7/99&91.8/98\\
 \enddata
\end{deluxetable}

%% file: nHCOTable1959LP.tex
\begin{deluxetable}{ccccccccc}
\tabletypesize{\scriptsize}
\tablecaption{Summary of LP spectral analysis for 1ES\,1959+650.  Observation ID details given in Table 3.  The Galactic N$_{\rm HI}$ column density was fixed to 1$\times10^{21}$cm$^{-2}$, as measured by \cite{kalberla} with a $\sim$3\% error. Observation ID 00035025004 is shown in bold and represents the best fit values for the models shown in the bottom panel of Figure 1.}
\tablewidth{0pt}
\tablehead{
\colhead{Observation}&    \colhead{Fixed} &  \colhead{Fixed}&  \colhead{Free}&  \colhead{Free}&   \colhead{Free}&  \colhead{Spectral}& \colhead{Fixed}& \colhead{Free}\\
\colhead{ID}&    \colhead{LP} &  \colhead{LP}&  \colhead{LP N$_{\rm HI}$}&  \colhead{LP}&    \colhead{LP}&   \colhead{Counts}& \colhead{LP}& \colhead{LP}\\
\colhead{}&   \colhead{$\alpha$} &  \colhead{$\beta$}&  \colhead{[$\times10^{22}$cm$^{-2}$]}&  \colhead{$\alpha$}&  \colhead{$\beta$}&  \colhead{}& \colhead{$\chi^2$/dof}& \colhead{$\chi^2$/dof}\\  
  
}
 \startdata
00035025001&1.83$\pm$0.02&0.68$\pm$0.03&0.15$\pm$0.01&2.16$\pm$0.07&0.29$\pm$0.086&32782&390.1/329&363.8/328\\
00035025002&1.73$\pm$0.03&0.69$\pm$0.06&0.18$\pm$0.03&2.2$\pm$0.1&0.1$\pm$0.2&9270&239.7/191&222.1/190\\
00035025003&1.69$\pm$0.03&0.62$\pm$0.05&0.14$\pm$0.02&1.9$\pm$0.1&0.4$\pm$0.1&14911&274.6/250&264.1/249\\
\textbf{00035025004}&\textbf{1.51$\pm$0.03}&\textbf{0.42$\pm$0.03}&\textbf{0.17$\pm$0.01}&\textbf{2.09$\pm$0.06}&\textbf{0.06$\pm$0.07}&\textbf{44642}&\textbf{537.6/416}&\textbf{481.1/415}\\
00035025005&1.59$\pm$0.02&0.63$\pm$0.04&0.17$\pm$0.01&1.98$\pm$0.08&0.21$\pm$0.09&26758&403.5/327&378.9/326\\
00035025006&1.67$\pm$0.02&0.62$\pm$0.03&0.19$\pm$0.01&2.16$\pm$0.06&0.07$\pm$0.07&44306&553.6/400&482.8/399\\
00035025007&1.72$\pm$0.03&0.57$\pm$0.03&0.18$\pm$0.01&2.19$\pm$0.06&0.04$\pm$0.07&43562&549.0/395&480.1/394\\
00035025008&1.74$\pm$0.02&0.79$\pm$0.03&0.19$\pm$0.01&2.29$\pm$0.07&0.17$\pm$0.08&36841&519.9/342&453.9/341\\
00035025009&1.75$\pm$0.02&0.75$\pm$0.03&0.20$\pm$0.01&2.36$\pm$0.07&0.06$\pm$0.08&35858&503.1/343&428.8/342\\
00035025010&1.83$\pm$0.02&0.71$\pm$0.04&0.19$\pm$0.01&2.36$\pm$0.08&0.1$\pm$0.1&26439&422.8/243&382.7/242\\
00035025016&1.5$\pm$0.2&1.2$\pm$0.3&0.1$\pm$0.2&1.48$\pm$0.9&1.15$\pm$1.0&1268&38.8/36&38.8/35\\
00035025027&2.00$\pm$0.04&0.67$\pm$0.08&0.16$\pm$0.03&2.4$\pm$0.2&0.2$\pm$0.2&6399&169.2/149&164.7/148\\
00035025028&2.18$\pm$0.04&0.39$\pm$0.08&0.17$\pm$0.03&2.6$\pm$0.2&-0.1$\pm$0.2&4749&121.6/119&116.0/118\\
00035025032&1.99$\pm$0.03&0.42$\pm$0.06&0.15$\pm$0.02&2.3$\pm$0.1&0.1$\pm$0.2&9745&172.0/194&165.9/193\\
00035025034&1.89$\pm$0.03&0.48$\pm$0.05&0.15$\pm$0.02&2.2$\pm$0.1&0.1$\pm$0.1&12162&217.0/218&208.9/217\\
00035025037&2.09$\pm$0.05&0.57$\pm$0.1&0.20$\pm$0.04&2.7$\pm$0.3&-0.2$\pm$0.3&3331&92.6/83&87.3/82\\
00035025038&1.91$\pm$0.05&0.63$\pm$0.09&0.15$\pm$0.04&2.2$\pm$0.2&0.3$\pm$0.2&4687&117.2/119&115.1/118\\
00035025041&2.05$\pm$0.04&0.27$\pm$0.08&0.17$\pm$0.03&2.5$\pm$0.2&-0.2$\pm$0.2&5488&129.6/138&123.3/137\\
00035025042&1.88$\pm$0.04&0.67$\pm$0.07&0.20$\pm$0.03&2.5$\pm$0.2&-0.01$\pm$0.2&7914&194.2/174&177.0/173\\
00035025043&2.10$\pm$0.04&0.45$\pm$0.09&0.13$\pm$0.03&2.3$\pm$0.2&0.2$\pm$0.2&4693&106.2/119&105.2/118\\
00035025044&2.05$\pm$0.02&0.51$\pm$0.03&0.11$\pm$0.01&2.12$\pm$0.07&0.43$\pm$0.08&30252&296.1/302&295.1/301\\
00035025045&1.85$\pm$0.05&0.47$\pm$0.08&0.07$\pm$0.03&1.7$\pm$0.2&0.64$\pm$0.2&4836&143.3/121&142.5/120\\
00035025046&1.84$\pm$0.04&0.38$\pm$0.07&0.14$\pm$0.03&2.1$\pm$0.2&0.1$\pm$0.2&6433&130.7/158&128.6/157\\
00035025047&1.91$\pm$0.05&0.323$\pm$0.08&0.05$\pm$0.02&1.6$\pm$0.2&0.7$\pm$0.2&5188&105.8/131&101.6/130\\
00035025048&1.87$\pm$0.05&0.42$\pm$0.09&0.14$\pm$0.03&2.1$\pm$0.2&0.2$\pm$0.2&5081&95.0/129&93.7/128\\
00035025049&1.68$\pm$0.06&0.69$\pm$0.1&0.13$\pm$0.04&1.9$\pm$0.2&0.5$\pm$0.2&4252&92.5/106&91.7/105\\
00035025050&1.67$\pm$0.07&0.93$\pm$0.1&0.19$\pm$0.05&2.2$\pm$0.3&0.4$\pm$0.3&4026&101.5/99&96.3/98\\
00035025051&1.88$\pm$0.07&0.63$\pm$0.1&0.17$\pm$0.05&2.3$\pm$0.3&0.2$\pm$0.3&3326&65.3/81&62.5/80\\
00035025052&2.00$\pm$0.05&0.34$\pm$0.09&0.16$\pm$0.03&2.4$\pm$0.2&0.0$\pm$0.2&4708&165.1/120&161.1/119\\
00035025053&1.68$\pm$0.06&0.68$\pm$0.09&0.11$\pm$0.03&1.8$\pm$0.2&0.6$\pm$0.2&5703&158.3/143&158.1/142\\
00035025054&1.81$\pm$0.06&0.7$\pm$0.1&0.13$\pm$0.04&2.0$\pm$0.2&0.5$\pm$0.2&4644&101.0/117&100.2/116\\
00035025056&2.00$\pm$0.06&0.5$\pm$0.1&0.14$\pm$0.04&2.2$\pm$0.3&0.2$\pm$0.3&3273&105.4/82&104.5/81\\
00035025057&2.09$\pm$0.06&0.7$\pm$0.1&0.18$\pm$0.04&2.6$\pm$0.3&0.1$\pm$0.3&3628&78.1/92&73.6/91\\
00035025058&1.90$\pm$0.04&0.66$\pm$0.08&0.16$\pm$0.03&2.3$\pm$0.2&0.2$\pm$0.2&6132&146.8/146&141.7/145\\
00035025059&1.76$\pm$0.04&0.55$\pm$0.06&0.15$\pm$0.02&2.1$\pm$0.1&0.2$\pm$0.2&9237&210.1/192&204.2/191\\
00035025062&2.34$\pm$0.05&0.5$\pm$0.1&0.07$\pm$0.03&2.1$\pm$0.2&0.7$\pm$0.3&3247&87.5/80&86.7/79\\
00035025063&2.30$\pm$0.07&0.5$\pm$0.1&0.16$\pm$0.05&2.7$\pm$0.3&0.1$\pm$0.4&2438&74.4/65&72.8/64\\
00035025064&2.06$\pm$0.07&0.5$\pm$0.1&0.07$\pm$0.03&1.9$\pm$0.2&0.7$\pm$0.3&3361&84.7/82&84.1/81\\
00035025066&1.75$\pm$0.07&0.7$\pm$0.1&0.14$\pm$0.05&2.0$\pm$0.2&0.4$\pm$0.3&3804&86.8/93&86.0/92\\
00035025067&1.88$\pm$0.06&0.8$\pm$0.1&0.10$\pm$0.04&1.9$\pm$0.2&0.8$\pm$0.3&3625&88.3/92&88.3/91\\
00035025068&2.09$\pm$0.09&0.4$\pm$0.2&0.16$\pm$0.07&2.4$\pm$0.4&0$\pm$0.5&2001&55.8/52&54.9/51\\
00035025069&2.0$\pm$0.1&0.6$\pm$0.2&0.23$\pm$0.09&2.7$\pm$0.4&-0.2$\pm$0.5&2073&58.0/55&54.4/54\\
00035025071&2.3$\pm$0.1&0.2$\pm$0.2&0.12$\pm$0.07&2.4$\pm$0.4&0.1$\pm$0.6&1442&38.5/39&38.5/38\\
00035025072&1.9$\pm$0.1&1.0$\pm$0.2&0.24$\pm$0.09&2.6$\pm$0.4&0.2$\pm$0.5&2364&78.2/62&74.0/61\\
00035025074&1.68$\pm$0.04&0.83$\pm$0.07&0.11$\pm$0.02&1.8$\pm$0.1&0.7$\pm$0.1&11257&209.2/215&208.9/214\\
00035025075&2.25$\pm$0.04&0.53$\pm$0.09&0.13$\pm$0.03&2.4$\pm$0.2&0.3$\pm$0.3&5504&111.8/129&110.9/128\\
00035025076&2.16$\pm$0.03&0.54$\pm$0.05&0.11$\pm$0.02&2.2$\pm$0.1&0.5$\pm$0.1&11140&211.7/194&211.4/193\\
00035025077&2.16$\pm$0.04&0.41$\pm$0.08&0.15$\pm$0.03&2.4$\pm$0.2&0.1$\pm$0.2&6044&120.4/141&117.3/140\\
00035025078&2.1$\pm$0.2&0.4$\pm$0.2&0.11$\pm$0.03&2.1$\pm$0.2&0.4$\pm$0.2&5500&127.1/134&127.1/133\\
00035025079&2.02$\pm$0.07&0.4$\pm$0.1&0.09$\pm$0.04&1.9$\pm$0.2&0.5$\pm$0.3&3841&100.2/93&100.1/92\\
00035025080&1.75$\pm$0.04&0.76$\pm$0.07&0.13$\pm$0.03&1.9$\pm$0.2&0.6$\pm$0.2&9006&223.6/186&222.4/185\\
00035025081&2.07$\pm$0.05&0.51$\pm$0.08&0.08$\pm$0.03&1.9$\pm$0.2&0.7$\pm$0.2&5230&100.1/126&97.6/125\\
00035025082&2.20$\pm$0.07&0.5$\pm$0.1&0.13$\pm$0.04&2.4$\pm$0.3&0.2$\pm$0.3&3207&85.4/80&84.6/79\\
00035025083&1.85$\pm$0.05&0.65$\pm$0.09&0.15$\pm$0.03&2.1$\pm$0.2&0.3$\pm$0.2&6148&138.5/146&135.9/145\\
00035025084&1.98$\pm$0.05&0.53$\pm$0.08&0.10$\pm$0.03&2.0$\pm$0.2&0.5$\pm$0.2&7082&161.6/160&161.6/159\\
00035025085&2.24$\pm$0.05&0.4$\pm$0.1&0.10$\pm$0.03&2.2$\pm$0.2&0.4$\pm$0.3&4448&133.4/112&133.4/111\\
00035025086&1.81$\pm$0.04&0.55$\pm$0.08&0.08$\pm$0.02&1.7$\pm$0.2&0.7$\pm$0.2&6083&146.2/154&145.6/153\\
00035025087&1.96$\pm$0.06&0.4$\pm$0.1&0.11$\pm$0.03&2.1$\pm$0.2&0.3$\pm$0.2&3916&65.3/96&65.0/95\\
00035025088&2.00$\pm$0.06&0.6$\pm$0.1&0.16$\pm$0.04&2.4$\pm$0.2&0.2$\pm$0.3&4892&105.2/125&102.4/124\\
00035025089&2.39$\pm$0.05&0.2$\pm$0.1&0.03$\pm$0.03&1.9$\pm$0.2&0.8$\pm$0.3&3142&92.9/76&88.4/75\\
00035025091&2.13$\pm$0.06&0.6$\pm$0.1&0.10$\pm$0.04&2.2$\pm$0.2&0.5$\pm$0.3&3934&89.5/98&89.5/97\\
 \enddata
\end{deluxetable}